\documentclass[aip,rsi,reprint]{revtex4-2}

\usepackage{graphicx}
\usepackage{units}
\usepackage{amssymb}
\usepackage[T1]{fontenc}
\usepackage[colorlinks=true, linkcolor=cyan, citecolor=cyan, urlcolor=cyan]{hyperref}

\begin{document}

\title{A compact and versatile cryogenic probe station for quantum device testing}

\author{Mathieu de Kruijf}
\altaffiliation
{Current address: London Centre for Nanotechnology, University College London, 17-19 Gordon Street, London WC1H 0AH, United Kingdom}

\author{Simon Geyer}

\author{Toni Berger}
\affiliation{Department of Physics, University of Basel, Klingelbergstrasse 82, CH-4056 Basel, Switzerland}

\author{Matthias Mergenthaler}
\affiliation{IBM Research Europe-Z\"urich, S\"aumerstrasse 4, CH-8803 R\"uschlikon, Switzerland}

\author{Floris Braakman}

\author{Richard J. Warburton}
\affiliation{Department of Physics, University of Basel, Klingelbergstrasse 82, CH-4056 Basel, Switzerland}

\author{Andreas V. Kuhlmann}
\thanks{Author to whom correspondence should be addressed: andreas.kuhlmann@unibas.ch}
\affiliation{Department of Physics, University of Basel, Klingelbergstrasse 82, CH-4056 Basel, Switzerland}

\date{\today}

\begin{abstract}
Fast feedback from cryogenic electrical characterization measurements is key for the development of scalable quantum computing technology. At room temperature, high-throughput device testing is accomplished with a probe-based solution, where electrical probes are repeatedly positioned onto devices for acquiring statistical data. In this work we present a probe station that can be operated from room temperature down to below \unit[2]{K}. Its small size makes it compatible with standard cryogenic measurement setups with a magnet. A large variety of electronic devices can be tested. Here, we demonstrate the performance of the prober by characterizing silicon fin field-effect transistors as a host for quantum dot spin qubits. Such a tool can massively accelerate the design-fabrication-measurement cycle and provide important feedback for process optimization towards building scalable quantum circuits. 
\end{abstract}

\maketitle

The successful development of solid-state quantum hardware, such as semiconductor- or superconductor-based qubits\cite{Loss1998,Takeda2022,Noiri2022,Xue2022,Philips2022,Mills2022,Wallraff2004,Arute2019,Krinner2022,Jafferis2022}, requires a close interplay between device design, fabrication, and measurement. Since these structures must typically be operated in a cryogenic environment, high-throughput device testing at low temperature (LT) is essential to accelerate the design-fabrication-measurement cycle. However, it often takes several days to obtain characterization data from cryogenic measurements when performed on just a few wire-bonded devices, and one cooldown does not provide a statistically meaningful sample size. Additionally, devices sensitive to electrostatic discharge can be damaged during wire bonding. A non-invasive solution for high-volume cryogenic testing is to adapt conventional room temperature (RT) wafer-scale probing to LTs. Indeed, the first cryogenic \unit[300]{mm}-wafer prober built to guide the industrial development of quantum devices has recently been released\cite{Pillarisetty2019}. While this probe system operates at measurement temperatures below \unit[2]{K}, it may not be suitable for academic research and small-scale prototyping. Furthermore, the integration of a magnet is challenging for such large wafer sizes, but would significantly strengthen the characterization toolbox. Other smaller size, commercially available systems often suffer from device temperatures well above \unit[4]{K} or a small number of probes. 

We present here the setup and operation of a cryogenic probe station allowing devices to be characterized at temperatures below $\unit[2]{K}$. It is designed for $2\times\unit[2]{cm^2}$ chips that are moved with respect to a multi-contact probe card using closed-loop piezo-based positioners. This prober is compact enough to fit inside a standard cryogenic magnet system, and is compatible with both direct-current (dc) and radio-frequency (rf) signals, therefore making it a versatile tool perfectly suited for research and prototyping. To showcase the benefit of this probe station for accelerating the design-fabrication-measurement cycle of cryogenic electronic devices, we characterize almost 50 silicon (Si) fin field-effect transistors (FinFETs) within one cooldown. 

At LTs, these FinFETs host quantum dots (QDs) \cite{Hanson2007,Kuhlmann2018,Geyer2021} such that we can obtain statistics on both transistor and QD properties. Spin qubits in Si-based QDs \cite{Loss1998,Zwanenburg2013,Stano2022,Burkard2021} rank among the prime candidates for implementing large-scale quantum processors since single- and two-qubit gate fidelities exceed the fault-tolerance threshold \cite{Noiri2022,Xue2022,Mills2022} and their similarities with respect to conventional transistors allow advanced industrial manufacturing processes to be exploited\cite{Zwerver2022,Elsayed2022,GonzalezZalba2021}. To take full advantage of the industry expertise to drive spin qubit development, cryogenic device testing must keep pace.

\begin{figure*}
\centering \includegraphics[width=\textwidth]{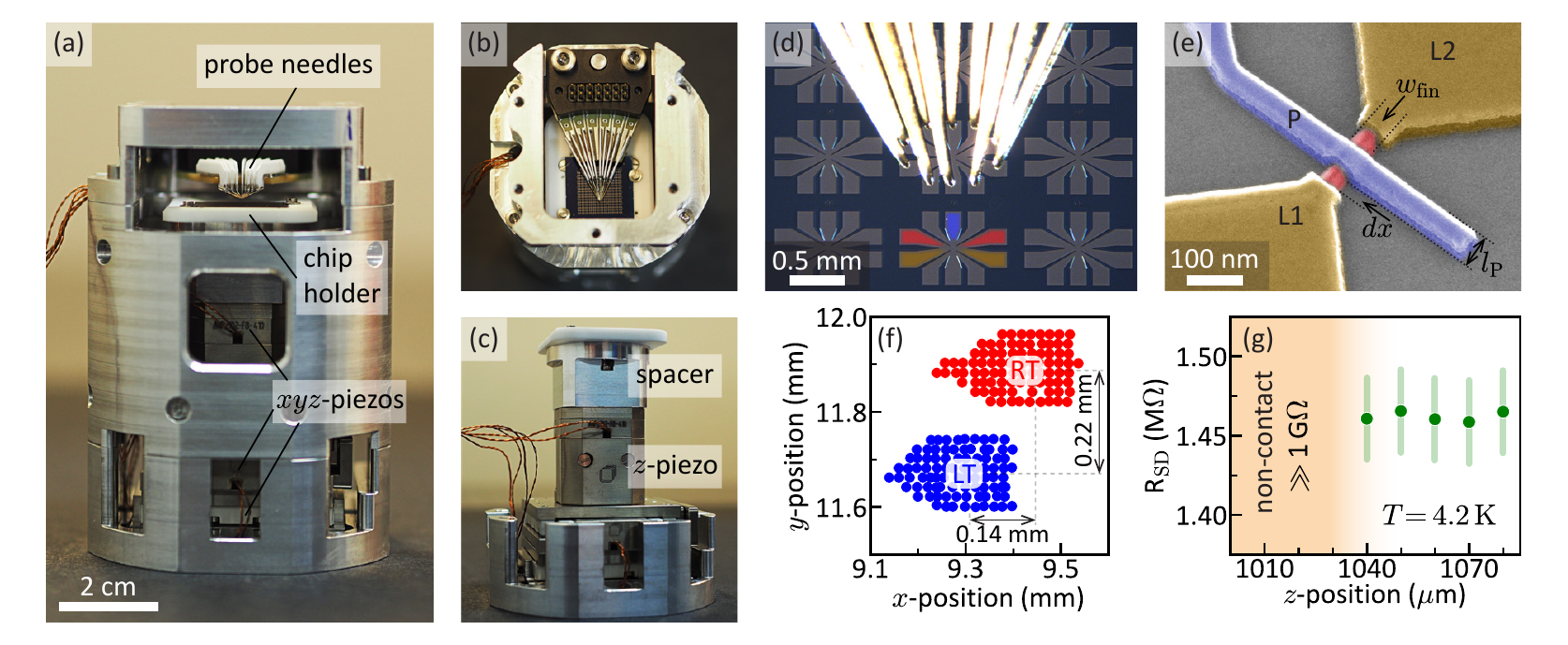}
\caption{\textbf{Cryogenic prober setup and devices under test}. Side-view (a),(c) and top-view (b) photographs of the prober that consists of an xyz piezo-based positioning unit, a holder for $2\times\unit[2]{cm^2}$ chips, and a multi-contact wedge probe card. In (c) the housing was partially removed in order to show the z-piezo. (d) Microscope image of the probe tips contacting a test device structure. (e) False-color scanning electron microscope image of a FinFET QD device showing the two lead gates L1 and L2 (yellow), and the plunger gate P (blue) wrapped around the Si fin (red). The 5 terminals required to contact such a device are highlighted in (d) using the same color code. (f) $xy$-marker scan, where red and blue points indicate a current flow between two adjacent needles exceeding \unit[0.1]{nA}. (g) Source-to-drain channel resistance as a function of the chip's $z$-position. Error bars correspond to $1\sigma$.}
\label{fig1}
\end{figure*}
The main idea of RT device probing on chip- or wafer-scale is to position probe needles in contact with the bond pads of a device, to run the measurements, and to repeat this procedure in an automated way for all devices to be tested. These needles can be assembled into an array on a probe card, which therefore serves as a tailored interface between the measurement hardware and the device. For aligning the probe needles, a camera is used for imaging. We adopt the same principle at LT but without imaging.

Photographs of our cryo-prober are presented in Figs.\ \ref{fig1}(a-c). The sample holder is mounted on top of an xyz piezo-based positioning unit, enabling a precise motion of the chip relative to the needles of a multi-contact wedge probe manufactured by GGB Industries. The housing of the prober is made of non-magnetic aluminum and can either be placed under a microscope at room temperature or attached via its top plate to the \unit[1]{K}-pot sample mount of a variable temperature insert (VTI). A major advantage of the prober's compact design is that it can be used with cryostats readily available in academic research labs. Horizontal motion in $x$- and $y$-direction is achieved using two attocube ANPx341/RES/LT closed-loop nanopositioners with \unit[20]{mm} travel range, which defines an upper limit for the chip area that can be tested. In addition, an attocube ANPz102/RES/LT drive with \unit[5]{mm} travel range is used for movements in $z$-direction. The ANPz102 has a maximum load of \unit[2]{N}, which is sufficient to push the chip into contact with the spring-loaded beryllium copper probe tips at cryogenic temperatures. This xyz-unit can be complemented by an attocube ANR101/RES/LT rotator. However, using sample holders with precisely machined cavities that host the chips, this degree of freedom was not required and thus replaced by a spacer [Fig.\ \ref{fig1}(c)]. All the positioners are compatible with both mK-temperatures and large magnetic fields. The chip's coordinates are monitored using the integrated resistive encoder of the drives, which however is temperature-sensitive, such that readout changes during cooldown or due to local heating of the resistor during the slip-stick movement of the positioners. 

We measure square-shaped chips with an edge length of \unit[2]{cm} and an area of ${1.5}\times\unit[1.5]{cm^2}$ occupied by 192 devices. The sample holder is made of either electrically conductive copper or insulating ceramic. The latter is seen in Figs.\ \ref{fig1}(a-c) and allows us to insulate the devices from the piezo positioners at RT, where the intrinsic-Si substrate is still slightly conductive. At LT the substrate's residual conductivity freezes out and a copper sample holder is used, thermally anchored to the VTI's \unit[1]{K}-pot, where the temperature $T$ is measured. 

The probe card used is a compact (max. length \unit[44]{mm}, max. width \unit[30]{mm}) multi-contact wedge with 10 dc needles that are arranged to match the devices' bond pad layout [Fig.\ \ref{fig1}(d)]. In addition to the dc probes, such a wedge can be equipped with probes specifically designed for rf-testing. 
\begin{figure*}
\centering 
\includegraphics[width=\textwidth]{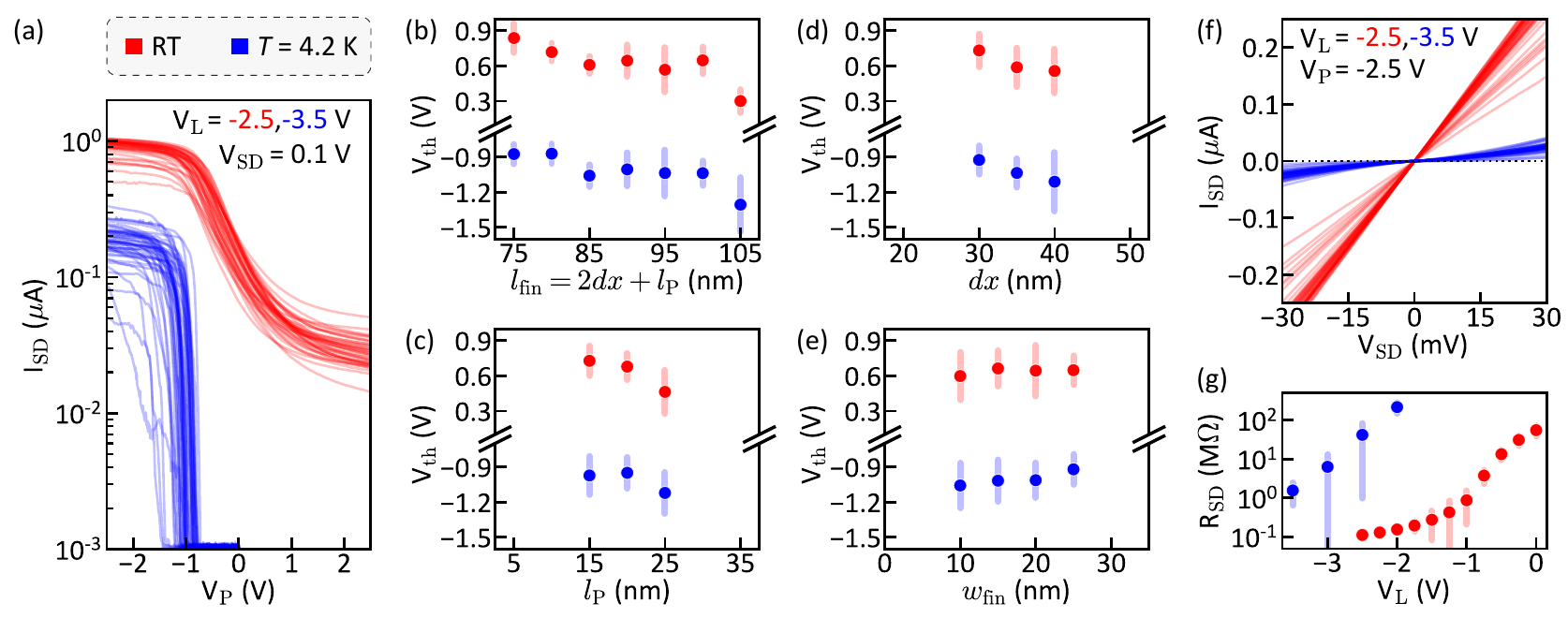}
\caption{\textbf{Transistor autoprobing at RT (red, 44 devices) and $\unit[{T}\,{=}\,{4.2}]{K}$ (blue, 48 devices)}. (a) Transistor turn-on curves in the high-bias regime. For the logarithmic current scale, zero-current flow from source to drain corresponds to \unit[1]{nA}. (b-e) Plunger gate threshold voltage dependence on the device dimensions. (f) I$_\mathrm{SD}$ versus V$_\mathrm{SD}$ curves measured in the transistor on-state, i.e.\ ${|\mathrm{V}_\mathrm{P}|}\,{>}\,{|\mathrm{V}_\mathrm{th}|}$. (g) Effect of lead gate voltage on the channel resistance. Plotted are the mean values with 1$\sigma$ error bars from all measured devices. }
\label{fig2}
\end{figure*}

We run the cryo-prober with a VTI and without optical access to the cold sample. Since the piezo's position readout depends on temperature and since in addition the materials contract during cooldown, the following procedure is applied to locate devices at LT. First, we determine the $z$-contact height $z_\mathrm{contact}$ by cooling down a $2\times\unit[2]{cm^2}$ Si chip with a metal (\unit[150]{nm} tungsten) surface coating. During cooldown the probe is centered in $x$- and $y$-directions over the chip and parked at the largest possible $z$-distance of $\unit[{\sim}{1.5}]{mm}$. With a small voltage applied to one needle, the cold chip is moved up in $z$-direction until this voltage is measured also on the other needles, indicating that all needles are in contact with the metal layer. Knowing $z_\mathrm{contact}$, the differences $\Delta x$, $\Delta y$ between room and low temperature ($x,y$)-coordinates are characterized using a marker structure on the chips to be tested. This marker can for example be the bond pads of the devices with all pads shorted together [Fig.\ \ref{fig1}(d)] or a malfunctioning device with leakage between two terminals. The latter approach is applied in Fig.\ \ref{fig1}(f) where a spatial map of the current between two probe tips, which are at slightly different voltages, is presented at RT and at $\unit[{T}\,{=}\,{4.2}]{K}$. From these measurements we extract ${\Delta x}\,{\simeq}\,\unit[140]{\mu m}$, ${\Delta y}\,{\simeq}\,\unit[220]{\mu m}$ and hence have knowledge of the cold device coordinates as the warm ones are known. From cooldown to cooldown $z_\mathrm{contact}$, $\Delta x$ and $\Delta y$ all vary by about $\unit[{\pm}{20}]{\mu m}$. Fig.\ \ref{fig1}(g) shows the source-to-drain resistance R$_\mathrm{SD}$ of a FinFET device (discussed later) at LT as a function of the chip's $z$-position: once $z_\mathrm{contact}$ is reached, R$_\mathrm{SD}$ is independent of $z$ and we therefore typically work at ${z}\,{=}\,{z_\mathrm{contact}}\,{+}\,\delta z$ with $\unit[{\delta z}\,{\simeq}\,{30}]{\mu m}$. When the device approach is automated for chip-scale testing, one needs to consider that the heat dissipated during the piezo's slip-stick motion can alter the sensor resistance and hence the position readout. This issue can be counteracted by achieving a good thermal anchoring of the positioners, and by dividing the movement into a coarse and fine step interleaved by a delay for thermalization. Another issue can arise from the positioner axes not being perfectly orthogonal lines, meaning that for example a step in $x$ also results in a small change in $y$. For ${150}\,{\times}\,\unit[150]{\mu m^2}$ bond pads and well-thermalized positioners, automated device approach was achieved in this work. We note that optical access to the cold sample would make the above described techniques to characterize $z_\mathrm{contact}$, $\Delta x$ and $\Delta y$ obsolete. Illuminating the chip with light, however, may alter the devices' charge noise environment\cite{Kuhlmann2013,Houel2012}.

We use this probe station to characterize a large number of Si FinFETs at both RT and LT. As shown in the scanning electron microscope image of Fig.\ \ref{fig1}(e), the devices under test have one plunger (P) and two lead gates (L1, L2); the p-type source and drain regions are made of platinum silicide (PtSi). These devices are designed to create accumulation-mode hole QDs; further details are given elsewhere\cite{Kuhlmann2018,Geyer2021}. Recently, we used similar devices to demonstrate single-qubit gate operations above \unit[4]{K}\cite{Camenzind2022} and two-qubit logic with anisotropic exchange\cite{Geyer2022} for holes in FinFETs. The chip's 192 devices are arranged on a 12-by-16 grid, in which the fin width $w_\mathrm{fin}$, plunger gate length $l_\mathrm{P}$, and gate spacing $dx$ [Fig.\ \ref{fig1}(e)] are varied (see supplementary material S1 for a map of the chip with the device dimensions). Unfortunately, the sample space diameter of the VTI available to us restricted the travel range of the $x$- and $y$-positioners such that only 60 devices could be approached. For testing of these five-terminal devices, 5 out of 10 needles land on a bond pad [Fig.\ \ref{fig1}(d)], the others on a \unit[{$\sim$}\,{100}]{nm}-thick silicon oxide layer. We test the devices by applying dc voltages and measuring the current response using a low-noise voltage source (BasPI SP927) complemented by a current monitoring box (BasPI SP1046). If more gain variability is required for current measurements, current-to-voltage amplifiers (BasPI SP983c) are utilized. All voltage signals are digitized with a National Instruments USB-6363 data acquisition card.

First, we compare in Fig.\ \ref{fig2} transistor characteristics at RT and $\unit[{T}\,{=}\,{4.2}]{K}$. Out of the 60 accessible devices, 48 (44) worked fine at LT (RT) (4 were damaged during cryogenic testing requiring larger applied voltages). We record the source-to-drain current I$_\mathrm{SD}$ while monitoring the gate leakage currents. If one of the latter exceeds a limit, the ongoing measurement is aborted and the tool automatically moves to the next device. In Fig.\ \ref{fig2}(a) the plunger gate voltage V$_\mathrm{P}$ is swept for a source-drain voltage V$_\mathrm{SD}$ of \unit[100]{mV} and a lead gate voltage V$_\mathrm{L}$ that is sufficiently above threshold (see supplementary material S2 for lead gate sweeps). While at LT the channel can be switched off completely, at RT a finite current flows in the transistor off-state due to conduction via the Si substrate. Moreover, the switching to the on-state is much steeper at \unit[4.2]{K} and occurs for a threshold voltage V$_\mathrm{th}$ of $\unit[{-1.01}\,{\pm}\,{0.17}]{V}$ (averaged over all devices) in contrast to $\unit[{+0.64}\,{\pm}\,{0.18}]{V}$ at RT. The dependence of V$_\mathrm{th}$ on the device dimensions is presented in Figs.\ \ref{fig2}(b-e). These curves highlight how probing of a statistically meaningful sample size enables trends to be identified. For instance, the threshold voltage decreases (increases) when the transistor channel length (width) is increased. Even though the V$_\mathrm{th}$-values are offset, the absolute change is almost the same at LT and RT. The plunger gate sweeps also reveal that the on-state current of the transistors is reduced at \unit[4.2]{K}, which is confirmed by the I$_\mathrm{SD}$-V$_\mathrm{SD}$ curves presented in Fig.~\ref{fig2}(f). While they are linear at RT, they are non-linear at LT where the presence of the Schottky barrier at the PtSi-Si junction is noticed more\cite{Kuhlmann2018}. As shown in Fig.\ \ref{fig2}(g), R$_\mathrm{SD}$ depends strongly on V$_\mathrm{L}$. While the resistance saturates at $\unit[{\sim}\,{100}]{k\Omega}$ for RT, no saturation effect is observed even at $\unit[{\mathrm{V}_\mathrm{L}}\,{=}\,{-3.5}]{V}$ for $\unit[{T}\,{=}\,{4.2}]{K}$. More negative lead gate voltages were not applied in order to avoid damaging the devices.

\begin{figure}
\centering 
\includegraphics[width=\columnwidth]{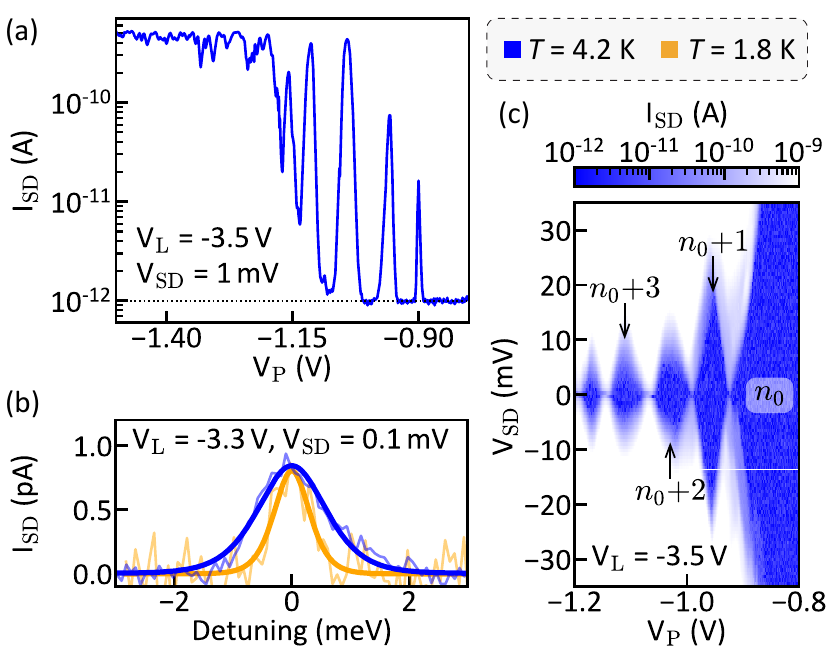}
\caption{\textbf{QD measurements at \unit[4.2]{K} (blue) and at \unit[1.8]{K} (orange)}. (a) Plunger gate sweep in the low bias regime with well-pronounced Coulomb blockade oscillations. (b) Extraction of the hole temperature by fitting the first observable Coulomb blockade peak. A hole temperature of $\unit[4.5\pm0.3]{K}$ and $\unit[2.3\pm0.4]{K}$ is found, respectively. (c) Charge stability diagram revealing the formation of a single QD. The data presented in this figure is measured on a device with ${w_\mathrm{fin}}\,\unit[{\sim}\,15]{nm}$, ${l_\mathrm{P}}\,\unit[{\sim}\,15]{nm}$, and ${dx}\,\unit[{\sim}\,35]{nm}$. }
\label{fig3}
\end{figure}
We next turn to the characterization of QDs, which requires cryogenic temperatures. An example for a plunger gate sweep recorded at $\unit[{T}\,{=}\,{4.2}]{K}$ with V$_\mathrm{SD}=\unit[1]{mV}$ and V$_\mathrm{L}=\unit[-3.5]{V}$ is given in Fig.\ \ref{fig3}(a) (see the supplementary material S3 for the same measurement repeated on all other devices). Here, a series of Coulomb resonances demonstrates single-hole tunneling via a QD that has formed beneath the plunger gate. Between the peaks, the device is in Coulomb blockade, that is, the number of holes residing on the QD is fixed\cite{Hanson2007}. Eventually the plunger gate's fringe fields lower the tunnel barriers such that for ${\mathrm{V}_\mathrm{P}}\,\unit[{\lesssim}\,{-1.3}]{V}$ a conducting channel opens\cite{Kuhlmann2018}. The closing of the Coulomb diamonds in Fig.\ \ref{fig3}(c) indicates the formation of a single QD below the plunger gate with charging energies up to $\unit[{\simeq}\,{20}]{meV}$. We examine the actual device cooling by means of Coulomb blockade thermometry: in the weak-coupling limit $\hbar\Gamma\ll k_B T$ and for small source-drain voltages the current resonances are thermally broadened, thus allowing us to determine experimentally the hole temperature $T_h$. Here $\hbar$ denotes the reduced Planck constant, $k_B$ the Boltzmann constant and $\Gamma$ the tunnel coupling between the QD and reservoir states. In Fig.\ \ref{fig3}(b), a high-resolution zoom-in on the first observable Coulomb peak recorded at V$_\mathrm{SD}\,\unit[{\simeq}\,{0.1}]{mV}$ for $\unit[{T}\,{=}\,{4.2}]{K}$ and $\unit[{T}\,{=}\,{1.8}]{K}$ is shown. By fitting the function\cite{Beenakker1991} $\mathrm{I}_{1}\cosh^{-2}[\alpha(\mathrm{V}_\mathrm{P,c}-\mathrm{V}_{\mathrm{P}})/(2k_B T_\mathrm{h})]+\mathrm{I}_0$ to the data a $T_h$ of $\unit[4.5\pm0.3]{K}$ and $\unit[2.3\pm0.4]{K}$ is extracted. Here, V$_\mathrm{P,c}$ denotes the peak center position, and $\alpha\,\unit[{\simeq}\,0.32]{eV/V}$ the plunger gate lever arm that is obtained from the Coulomb diamond plotted in Fig.\ \ref{fig3}(c). The hole temperatures are slightly above the ones measured with the \unit[1]{K}-pot thermometer, indicating an insufficient thermal anchoring of the chip or the VTI's dc lines. Both can be improved and device temperatures down to $\unit[{\simeq}\,1.5]{K}$ are within reach. 

Finally, we remark that the cryo-prober allows currents to be measured with the same low noise level as achieved when characterizing wire-bonded devices; Fig.\ \ref{fig3} confirms that currents $\unit[{<}{100}]{fA}$ are detectable.

In conclusion, we designed, built and operated a cryogenic probe station enabling automated probing of electronic devices at temperatures below \unit[2]{K}. The instrument's value for accelerating the design-fabrication-measurement cycle is demonstrated by testing almost 50 Si FinFET QD devices at both room and low temperature. Probing of a statistically significant number of devices allows us to extract reliably transistor and QD properties, and to identify trends and patterns in the data. Besides high-throughput device testing, the prober can be utilized to search for the highest quality devices for subsequent dilution refrigerator experiments. The prober's degree of automation can be enhanced further by (i) implementing optical access to the cold sample enabling device localization via image pattern recognition, and (ii) using machine learning for completely automatic tuning of quantum devices\cite{Lennon2019,Moon2020,Severin2021}. Furthermore, the prober can be upgraded with a magnet and rf probes such that not only transistor and QD characteristics but also statistics on qubit parameters can be obtained. This functionality will be especially useful for Si QD spin qubits since recent work has shown that these can be operated at temperatures of up to \unit[5]{K}\cite{Yang2020,Petit2020,Camenzind2022,Petit2022}.\\ 

\newpage
\noindent\textbf{Supplementary material}\\
See supplementary material for a map of the chip with device dimensions, lead gate sweeps at RT, and low-bias plunger gate sweeps at \unit[4.2]{K} performed on all devices.\\

\noindent\textbf{Acknowledgments}\\
We thank M.\ Poggio and D.\ Zumb\"uhl for fruitful discussions. We acknowledge technical support at the University of Basel by S.\ Martin, M.\ Steinacher and R.\ Maffiolini, as well as support by the cleanroom operation team, particularly U.\ Drechsler, A.\ Olziersky and D.\ D.\ Pineda, at the IBM Binnig and Rohrer Nanotechnology Center. This work was partially supported by the NCCR SPIN, the Swiss Nanoscience Institute (SNI), the Georg H. Endress Foundation, and the EU H2020 European Microkelvin Platform EMP (grant nr.\ 824109).\\

\noindent\textbf{Data availability}\\
The data that support the findings of this study are available from the corresponding author upon reasonable request.

\end{document}


\title{Supplementary material to: ''A compact and versatile cryogenic probe station for quantum device testing``}

\author{Mathieu de Kruijf}
\altaffiliation
{Current address: London Centre for Nanotechnology, University College London, 17-19 Gordon Street, London WC1H 0AH, United Kingdom}

\author{Simon Geyer}

\author{Toni Berger}
\affiliation{Department of Physics, University of Basel, Klingelbergstrasse 82, CH-4056 Basel, Switzerland}

\author{Matthias Mergenthaler}
\affiliation{IBM Research Europe-Z\"urich, S\"aumerstrasse 4, CH-8803 R\"uschlikon, Switzerland}

\author{Floris Braakman}

\author{Richard J. Warburton}

\author{Andreas V. Kuhlmann}
\thanks{Author to whom correspondence should be addressed: andreas.kuhlmann@unibas.ch}
\affiliation{Department of Physics, University of Basel, Klingelbergstrasse 82, CH-4056 Basel, Switzerland}

\date{\today}

\begin{abstract}
In ''A compact and versatile cryogenic probe station for quantum device testing`` we implement a probe station allowing for high-throughput testing of cryogenic quantum devices. This prober is utilized to characterize a large number of silicon fin field-effect transistors, which at low temperature can host QD spin qubits. In this supplementary material we provide details on the device dimensions, and present room temperature lead gate sweeps and low temperature plunger gate sweeps for all devices. 
\end{abstract}

\maketitle

\section*{S1.\ Device dimensions}
\begin{figure}[h]
\centering 
\includegraphics[width=0.5\textwidth]{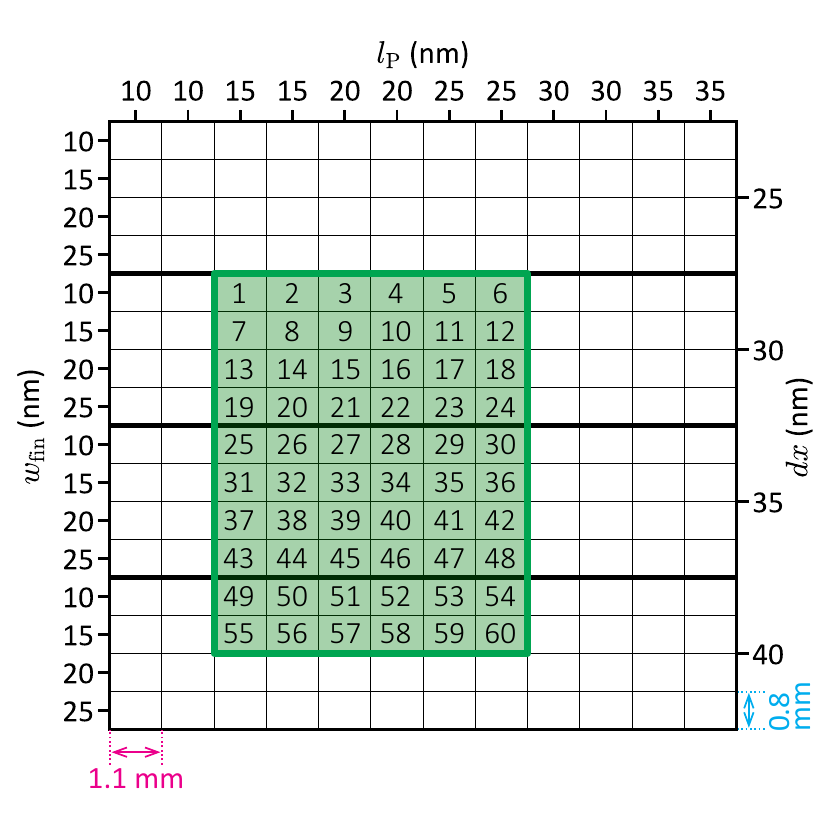}
\caption{We fabricated $2\times\unit[2]{cm^2}$ chips that have an active area of ${1.5}\times\unit[1.5]{cm^2}$ occupied by 192 devices. A single device extends over an area of ${1.1}\times\unit[0.8]{mm^2}$. Across the chip the fin width ($w_\mathrm{fin}=\unit[10, 15, 20, 25]{nm}$), plunger gate length ($l_\mathrm{P}=\unit[10, 15, 20, 25, 30, 35]{nm}$) an gate spacing ($dx=\unit[25, 30, 35, 40]{nm}$) are varied. There are always two adjacent devices that have the same dimensions. The VTI's sample space diameter of \unit[71]{mm} limited the travel range of the piezo positioners such that only the 60 devices within the green shaded region could be accessed. The data shown in Fig.\ 3 of the main article is obtained from device 7 with $w_\mathrm{fin}=\unit[15]{nm}$, $l_\mathrm{P}=\unit[15]{nm}$ and $dx=\unit[30]{nm}$.}
\label{figS1}
\end{figure}

\newpage
\section*{S2.\ Room temperature lead gate sweeps}

\begin{figure}[h]
\centering 
\includegraphics[width=\textwidth]{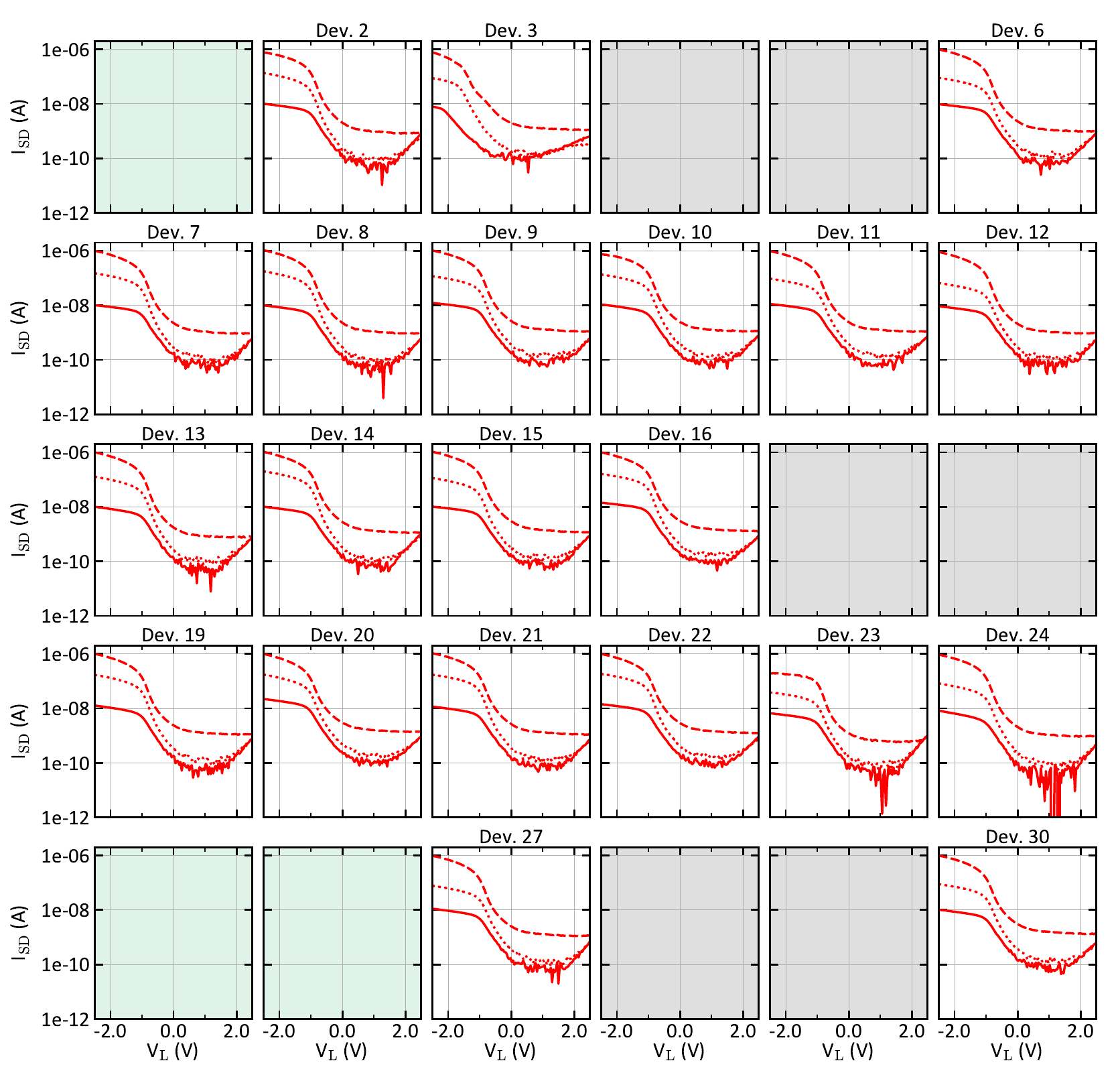}
\caption{Source-drain current I$_\mathrm{SD}$ as a function of lead gate voltage V$_\mathrm{L}$ for a source-drain voltage V$_\mathrm{SD}$ of \unit[100]{mV} and a plunger gate voltage V$_\mathrm{P}$ of \unit[+2.5]{V} (solid curve), \unit[0]{V} (dotted curve) and \unit[-2.5]{V} (dashed curve). 45 devices were auto-probed at room temperature. The empty panels indicate non-working devices, the green shaded ones broke during cryogenic testing. For the log-scale, zero-current flow from source to drain corresponds to \unit[1]{pA}.}
\label{figS2a}
\end{figure}

\newpage
\begin{figure}[h]
\centering 
\includegraphics[width=\textwidth]{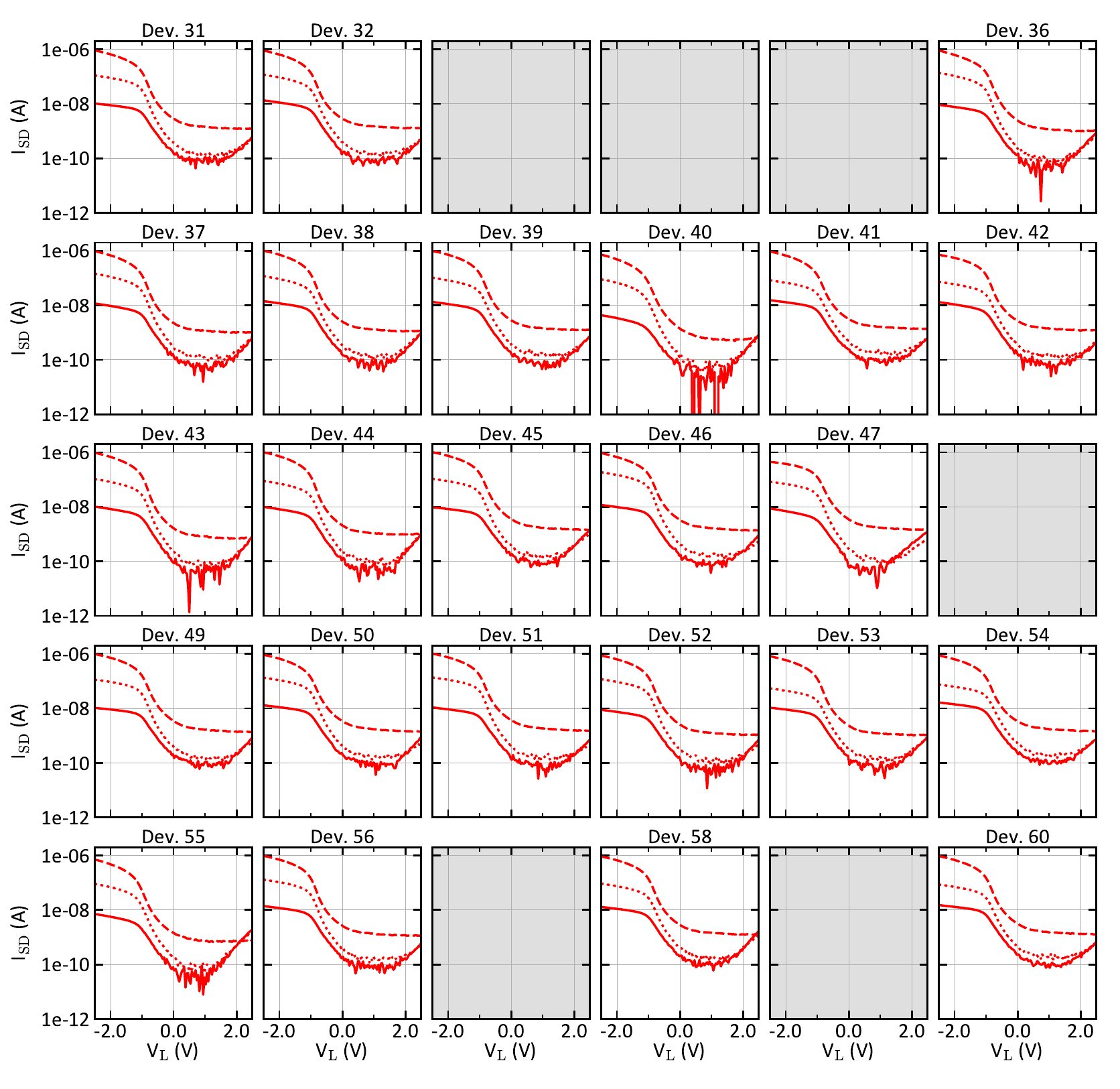}
\caption{Continuation of Fig.\ \ref{figS2a}}
\label{figS2b}
\end{figure}

\newpage
\section{Low temperature plunger gate sweep at V$_\mathrm{SD}=\unit[1]{mV}$}

\begin{figure}[h]
\centering 
\includegraphics[width=\textwidth]{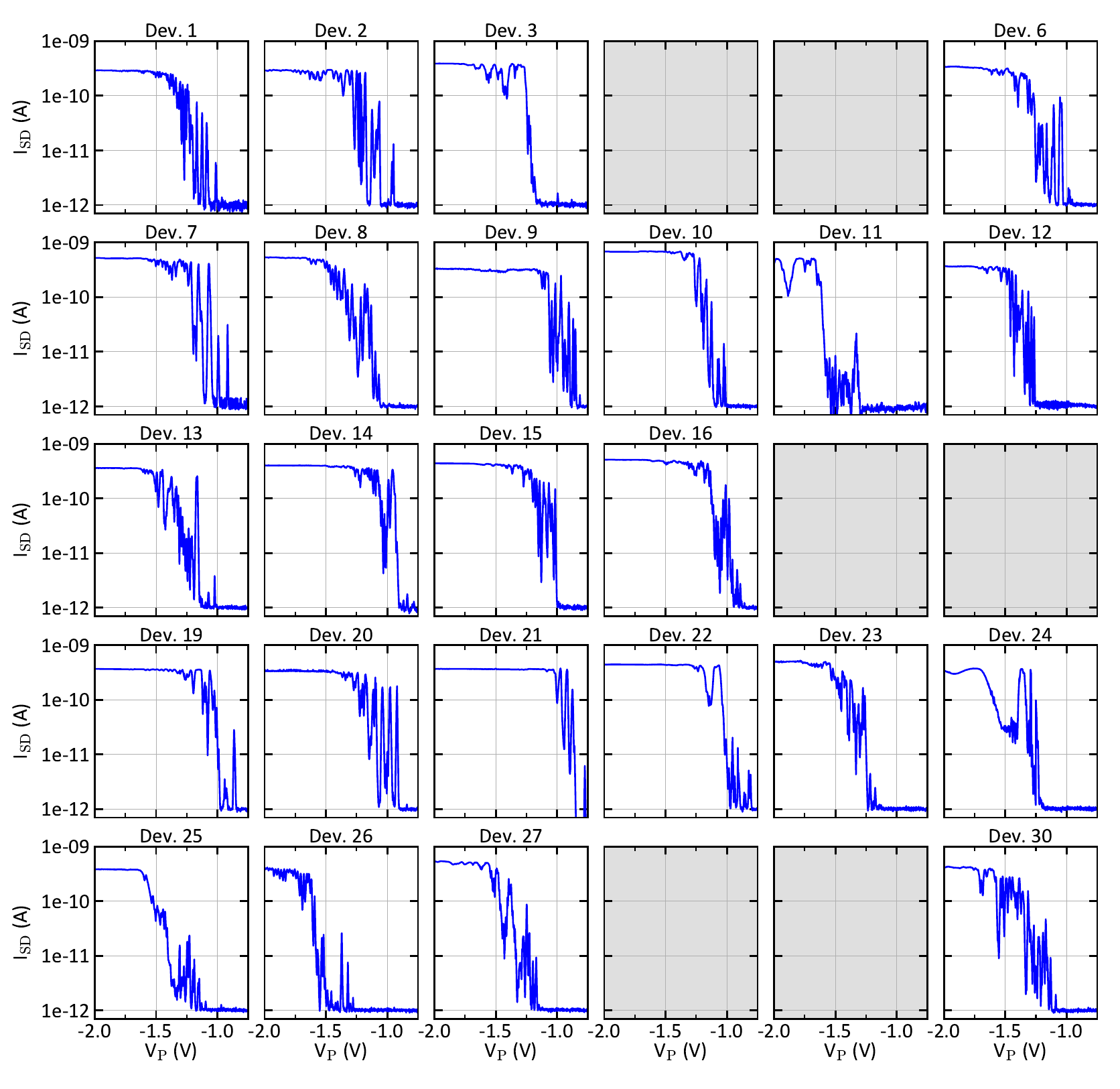}
\caption{Source-drain current I$_\mathrm{SD}$ as a function of plunger gate voltage V$_\mathrm{P}$ for a source-drain voltage V$_\mathrm{SD}$ of \unit[1]{mV}. 48 devices were auto-probed at \unit[4.2]{K}. The empty panels indicate non-working devices. For the log-scale, zero-current flow from source to drain corresponds to \unit[1]{pA}.}
\label{figS3a}
\end{figure}

\newpage
\begin{figure}[h]
\centering 
\includegraphics[width=\textwidth]{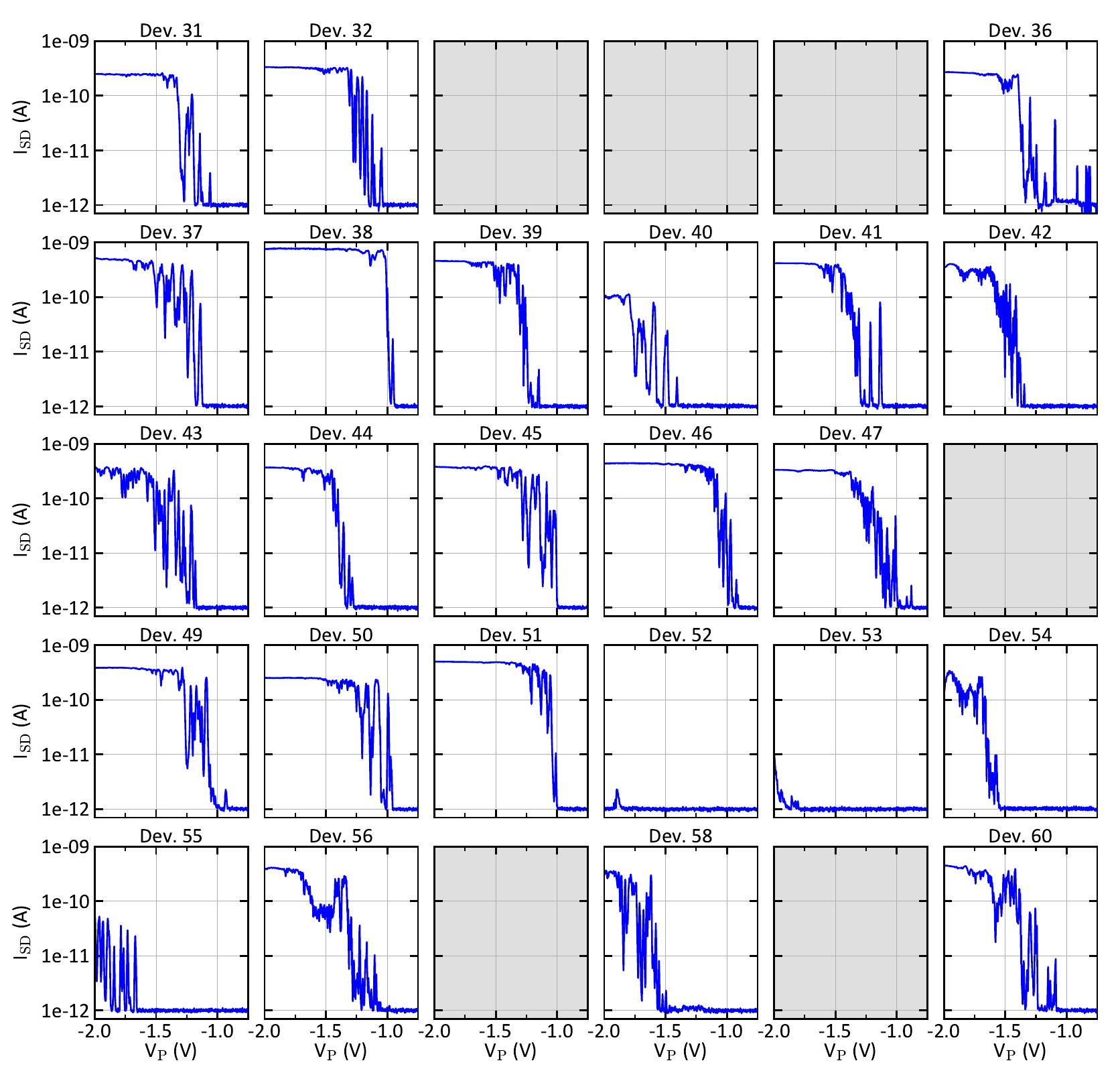}
\caption{Continuation of Fig.\ \ref{figS3a}}
\label{figS3b}
\end{figure}